\documentclass{article}
\usepackage{amssymb}
\usepackage{spconf,amsmath,graphicx}
\usepackage{color}
\usepackage{multirow}
\usepackage{booktabs}
\usepackage{graphicx}

\title{enhancing expressiveness in dance generation via integrating frequency and music style information}

\name{Author(s) Name(s)\thanks{Thanks to XYZ agency for funding.}}
\address{Author Affiliation(s)}
\name{
\begin{tabular}{c}
Qiaochu Huang$^{1,*,\ddagger}$, Xu He$^{1,*}$\thanks{$\ddagger$ Work conducted when the first author was an intern at XVerse Inc.}\thanks{* The authors contributed equally. \quad $\dagger$ Corresponding author.}, Boshi Tang$^1$, Haolin Zhuang$^1$,   Liyang Chen$^1$,\\ Shuochen Gao$^1$, \textit{Zhiyong Wu$^{1,3,\dagger}$, Haozhi Huang$^2$},\textit{ Helen Meng$^{3}$}
\end{tabular}
\vspace{-0.2cm}
}

\address{$^1$
    Shenzhen International Graduate School, Tsinghua University, Shenzhen, China \\
  $^2$ XVerse Inc., Shenzhen, China   
    $^3$ The Chinese University of Hong Kong, Hong Kong SAR, China\\
    \small{
        \{hqc22, hex22\}$@$mails.tsinghua.edu.cn, 
        zywu$@$sz.tsinghua.edu.cn,
        huanghz08@gmail.com
    }
}

\begin{document}
\ninept
\maketitle
\begin{abstract}
Dance generation, as a branch of human motion generation, has attracted increasing attention. Recently, a few works attempt to enhance dance expressiveness, which includes genre matching, beat alignment, and dance dynamics, from certain aspects. However, the enhancement is quite limited as they lack comprehensive consideration of the aforementioned three factors. In this paper, we propose ExpressiveBailando, a novel dance generation method designed to generate expressive dances, concurrently taking all three factors into account. Specifically, we mitigate the issue of speed homogenization by incorporating frequency information into VQ-VAE, thus improving dance dynamics. Additionally, we integrate music style information by extracting genre- and beat-related features with a pre-trained music model, hence achieving improvements in the other two factors. Extensive experimental results demonstrate that our proposed method can generate dances with high expressiveness and outperforms existing methods both qualitatively and quantitatively\footnote{Our project page https://github.com/thuhcsi/ExpressiveBailando includes demos and code.}.

\end{abstract}
\section{Introduction}
Music-driven dance generation has become a heated research topic recently, where dance expressiveness is crucial as it enhances the attractiveness and vividness of the generated dances.
While artificial intelligence (AI) choreographies have been widely used in such domains as choreography assistance~\cite{bailando} and dance teaching~\cite{danceteaching}, generating expressive dances using AI remains challenging.

In the field of choreography, genre matching, beat alignment, and dance dynamics are recognized as three essential factors contributing to dance expressiveness~\cite{gtn-bailando,beat,dynamics}. To be specific, genre matching and beat alignment describe the consistency between dance and music at different levels. The former focuses on the style while the latter concentrates on the rhythm, both aiming at providing a harmonious audio-visual experience. Dance dynamics refers to how dancers perform, specifically fast or slowly, explosively or continuously, concerned primarily with speed variation of the motions. Varying dance dynamics creates a strong visual impact and delivers intense emotions, thus impressing the audience.


Early works~\cite{learn2dance_earlyworks, musicsim_earlyworks, musiccontent_earlyworks} in dance generation often employ graph-based methods, which crop existing dance sequences into fixed-length fragments and reassemble them to form new dances based on carefully designed rules. 
While these methods guarantee the physical plausibility of the generated dances, they fail to create new dance motions and cannot match varying music rhythms for the fixed tempos of the cropped fragments.
Recently, deep neural networks have emerged as a promising method for dance generation, including CNNs~\cite{cnn_deep}, RNNs~\cite{rnn_groovenet, rnn_weak}, GANs~\cite{gan_deepdance}, Transformers~\cite{aist,transformer_danceformer}, and Diffusion models~\cite{edge}.
These approaches can generate diverse motions for music of various rhythms and styles. However, they lack consideration of dance expressiveness, leading to insufficient attractiveness and vividness of the generated dances.

On the other hand, some works make initial attempts to enhance dance expressiveness. 
Bailando~\cite{bailando} 
employs an actor-critic-based reinforcement learning scheme~\cite{actor} to improve beat alignment. 
GCDG~\cite{gcdg} incorporates genre embedding into transformer decoder to improve genre matching.
GTN-Bailando~\cite{gtn-bailando} introduces a genre token network, inferring genre information from music for better genre matching and consistency.
However, these studies only touch upon certain aspects of dance expressiveness without comprehensive consideration of the three factors mentioned above, thus having limited effects on enhancing dance expressiveness.

In this paper, we introduce a novel dance generation method, ExpressiveBailando, to enhance dance expressiveness with concurrent consideration for genre matching, beat alignment, and dance dynamics. 
First, we propose Frequency Complemented VQ-VAE (FreqVQ-VAE) to enhance dance dynamics by mitigating the issue of speed homogenization of the generated dances. More specifically, we utilize Frequency Complement Modules (FCM) to integrate frequency information into VQ-VAE~\cite{vqvae} under the guidance of Focal Frequency Loss (FFL)~\cite{ffl}.
Second, to enhance expressiveness in terms of genre matching and beat alignment, we leverage a pre-trained music model, MERT~\cite{mert}, which extracts music style information in the form of genre- and beat-related features.
Experimental results demonstrate our proposed method excels in both qualitative and quantitative evaluations, generating dances with higher expressiveness concerning genre matching, beat alignment, and dance dynamics.




\section{Method}
\label{sec:format}
\begin{figure*}[h]
    \centering
    \includegraphics[width=1\textwidth]{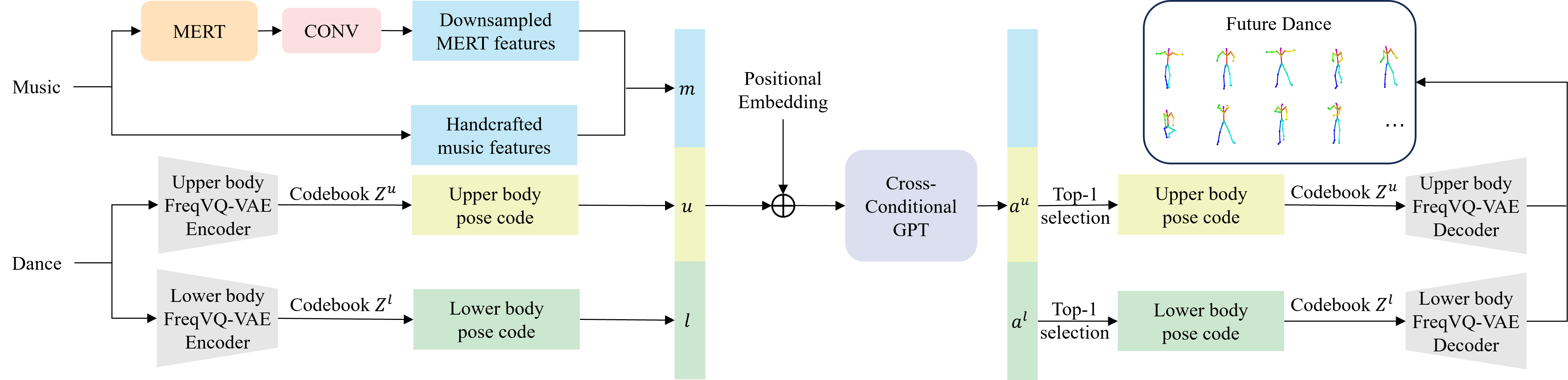} 
    \caption{The overall architecture for ExpressiveBailando. MERT features are extracted, downsampled, and concatenated with handcrafted music features to form the music conditional input to cross-conditional GPT. Future dances are generated by decoding the upper and lower body pose codes predicted by the GPT with the FreqVQ-VAE decoders.}
    \label{fig:expressivebailando}
    \vspace{-0.3cm}
\end{figure*}


The architecture of our proposed ExpressiveBailando is illustrated in Fig.~\ref{fig:expressivebailando}. We first train two FreqVQ-VAEs on upper and lower half bodies separately by reconstructing dance sequences. Subsequently, we employ a cross-conditional GPT~\cite{bailando} for dance generation. To be specific, we feed the dance into the FreqVQ-VAE encoders to obtain the upper and lower body pose codes and embed them. Then these embeddings $u$ and $l$ are concatenated with the music conditional input $m$, including downsampled MERT features and handcrafted music features, to form the input of the cross-conditional GPT. The GPT outputs the probabilities of the future upper and lower body pose codes, denoted as $a^u$ and $a^l$ respectively. We then select the most probable pose codes (top-1 selection) and decode them using the FreqVQ-VAE decoders to generate future dance. In inference, the GPT autoregressively predicts pose codes according to the initial pose code and the entire music, followed by FreqVQ-VAE decoders generating a new dance.
\subsection{Frequency Complemented VQ-VAE (FreqVQ-VAE)}

In previous dance generation works~\cite{bailando}, VQ-VAE consists of an encoder, a decoder, and a learnable discrete codebook $Z\in \mathbb{R}^{M\times d}$, where $M$ and $d$ stand for the size and dimension of the codebook respectively. At training time, VQ-VAE first maps a dance sequence $P \in \mathbb{R}^{T\times (J\times 3)}$ into latent features $e \in \mathbb{R}^{T_q \times d}$, quantizes them, and then reconstructs the dance sequence $\hat{P} \in \mathbb{R}^{T\times (J\times 3)}$ from the quantized features $e_q \in \mathbb{R}^{T_q \times d}$, where $T$ is the length of dance sequence and $J$ is the number of joints, $T_q = \frac{T}{r}$ and $r$ is the downsampling rate. After training, each entry in the codebook represents a meaningful dance pose, and the integer $p$ used to index each entry is called the pose code~\cite{bailando}. Thus, we can generate new dance sequences by decoding new combinations of pose codes.

\begin{figure}[t]
    \centering
    \includegraphics[width=1\linewidth]{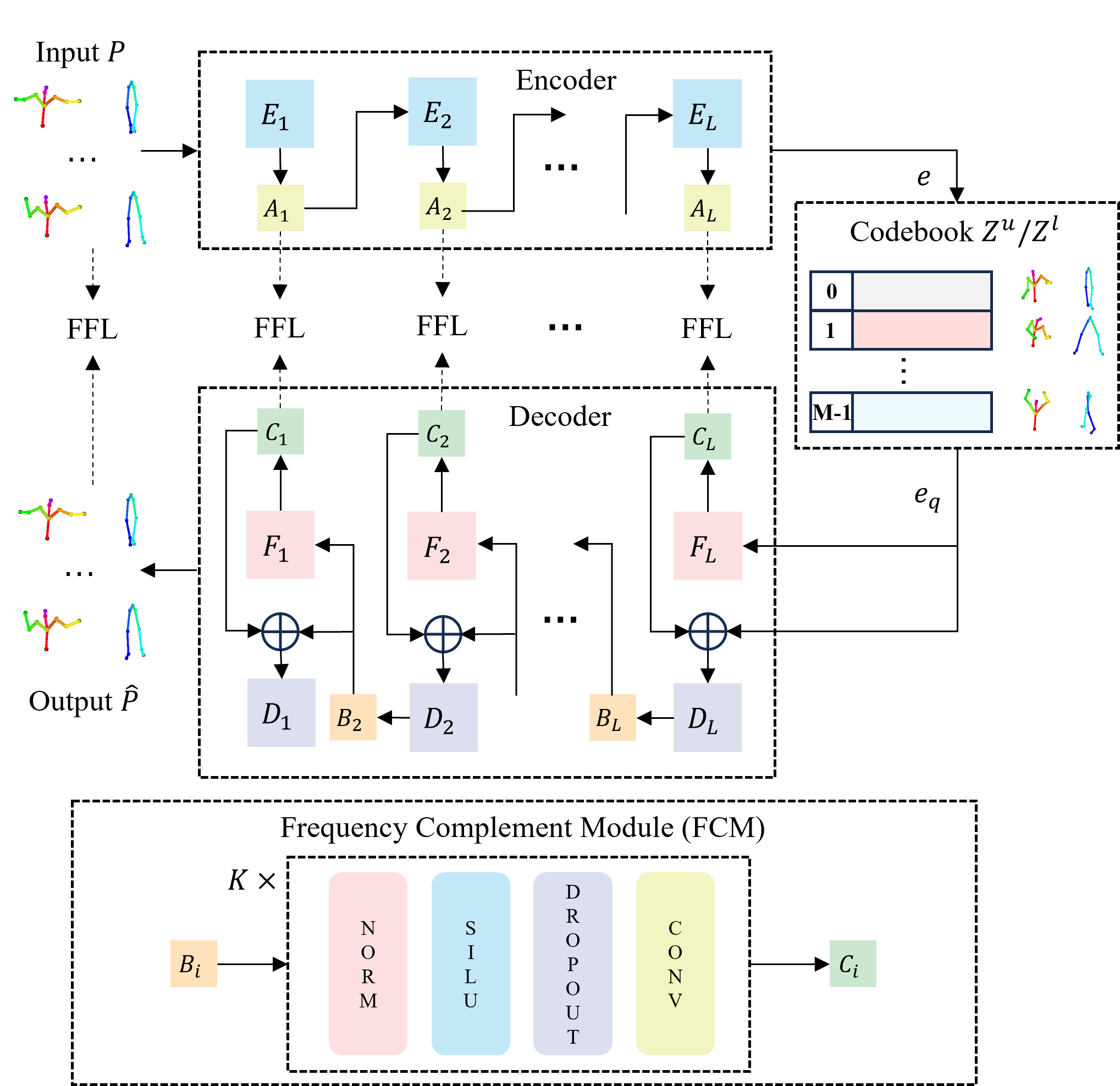} 
    \caption{Architecture of the proposed FreqVQ-VAE.}
    \label{fig:freqvavae}
    \vspace{-0.4cm}
\end{figure}

However, VQ-VAE reportedly suffers from a loss of frequency information in the decoding process~\cite{favae}, which leads to speed homogenization, i.e. lack of variation in speed, of the generated dances. This problem results in diminished dance dynamics. To mitigate this issue, inspired by~\cite{favae}, we propose FreqVQ-VAE with FFL guiding FCMs to integrate frequency information from each encoder layer into the output of the corresponding decoder layer. The architecture of FreqVQ-VAE is illustrated in Fig.~\ref{fig:freqvavae}.



Specifically, each FCM contains a stack of 1-D convolutional layers accompanied by activation layers. We incorporate FCMs into the VQ-VAE decoder through residual connections to complement the output of each decoder layer.
As depicted in Fig.~\ref{fig:freqvavae}, let $E_i$ and $D_i$ denote the $i$-th layer of the encoder and decoder, $A_i$ and $B_i$ be the output of $E_i$ and $D_i$. $F_i$ represents the $i$-th FCM to output $C_i$. The decoder of FreqVQ-VAE can be expressed as:
\begin{equation}
\begin{aligned}
    B_{L} &= D_{L}(e_q+C_{L}) = D_{L}(e_q+F_{L}(e_q)) \\ 
    B_{i} &= D_{i}(B_{i+1}+C_{i}) = D_{i}(B_{i+1}+F_{i}(B_{i+1})) \\
    i&=2,3,\cdots, L-1 \\
\end{aligned}
\end{equation}
where $L$ is the number of decoder layers,

Furthermore, we utilize FFL to guide FCMs to learn frequency information from the corresponding encoder layer output explicitly. We first conduct the 1-D Discrete Fourier Transform (DFT) on each channel of the $A_i$ and  $C_i$:
\begin{equation}
    X(k) = \sum_{n=0}^{N-1}x(n)\cdot e^{-j\frac{2\pi}{N}kn},
\end{equation}
where 
$N$ is the dimensionality of the feature, $x(n)$ is the value at the $n$-th position in each feature, and $X(k)$ is the value of the corresponding spectrum at frequency $k$. 
FFL is computed as follows:
\begin{equation}
    \mathrm{FFL}(A_i,C_i) = \frac{1}{N|C_i|}\sum_{c=0}^{|C_i|}\sum_{k=0}^{N-1}sg[w(k)]J(k),
\end{equation}
where $sg[\cdot]$ denotes the stop gradient function, $w(k)=|X_{A_i}(k)-X_{C_i}(k)|$ weighs losses from different frequency components according to frequency disparities, $J(k)=|X_{A_i}(k)-X_{C_i}(k)|^2$ is the frequency distance between $A_i$ and $C_i$, and $|C_i|$ denotes the number of $C_i$'s channels. FFL can be seen as a weighted average of the frequency distance between $A_i$ and $C_i$. In this way, it pays more attention to frequency components with a large distance to bridge frequency domain gaps~\cite{ffl}.

We also apply FFL between the input $P$ and the reconstructed dance sequence $\hat{P}$, and the total training loss for FreqVQ-VAE is formulated as:
\begin{equation}
\label{eq:lfreq}
    \mathcal{L}_{Freq} = \alpha_1 \mathrm{FFL}(P, \hat{P}) + \alpha_2 \sum_{i=0}^{L-1}\mathrm{FFL}(A_i, C_i) + \mathcal{L}_{VQ}
\end{equation}
where $\mathcal{L}_{VQ}$ is the original VQ-VAE training loss~\cite{bailando}, $\alpha_1,\ \alpha_2 \in [0,1]$ are hyperparameters. Following Bailando, we train two separate FreqVQ-VAEs for the upper and lower body so that the codebooks $Z^u$ and $Z^l$ can cover more dance poses.


\subsection{Music Features}

Most previous works~\cite{aist,bailando} in dance generation focus on the improvement of model architecture, but neglect the processing of music. Such studies take conventional handcrafted features, e.g., Mel Frequency Cepstrum Coefficient (MFCC), as the music inputs for dance generation models. EDGE~\cite{edge} utilizes Jukebox~\cite{jukebox} for the extraction of music features as music inputs and achieves better results. It exhibits the superiority of leveraging a pre-trained music model in dance generation.

Motivated by EDGE, we likewise employ a pre-trained music model for music feature extraction. Here we utilize MERT~\cite{mert}, for its competitive performance in various downstream Music Information Retrieval (MIR) tasks, including genre classification and beat tracking. The features extracted by MERT contain rich music style information concerning genre and beat, which helps to improve the genre matching and beat alignment of the generated dances. Specifically, we feed the music into MERT to obtain the output from its final layer. Then we downsample this output through 1-D convolutional layers 
and concatenate the downsampled MERT features with handcrafted features in the feature dimension to form the music conditional input $m$ for the cross-conditional GPT, as shown in Fig.~\ref{fig:expressivebailando}.

\section{EXPERIMENTS}
\label{sec:pagestyle}
\subsection{Dataset}
To thoroughly examine the effectiveness of our methods, we employ the largest publicly available 3D dance-music-aligned dataset, namely AIST++~\cite{aist}, for training and evaluation. AIST++ comprises 992 high-quality 60-FPS dance sequences in the SMPL format~\cite{smpl}, spanning 10 distinct dance genres. We utilize the train/test splits provided by the original dataset with 952 samples for training and 40 for testing.

\subsection{Experiment Setup}
For both the upper and lower body FreqVQ-VAEs, we set the codebook size $M$ and dimension $d$ to 512, with a downsampling rate $r$ of 8. Each FCM comprises two convolutional layers, with a dropout rate of 0.2. For Eq. \ref{eq:lfreq}, the parameters $\alpha_1$ and $\alpha_2$ are respectively set to 1 and 0.1. The training is conducted for 500 epochs using the Adam optimizer~\cite{adam} with $\beta_1=0.9,\beta_2=0.99$, and a learning rate of $3.0\times 10^{-5}$. For the cross-conditional GPT, we adopt the model configurations and training setup from Bailando~\cite{bailando}. We use the publicly available MERT-330M 
checkpoint and employ two convolutional layers with strides of 2 and 5, respectively, to downsample the MERT features. The handcrafted music features include MFCC, MFCC delta, constant-Q chromagram, tempogram, and onset strength.
All dance and music are cropped into 4-seconds clip for the training of FreqVQ-VAE and cross-conditional GPT.

We compare our proposed method with Bailando~\cite{bailando}, EDGE \cite{edge}, and GTN-Bailando~\cite{gtn-bailando}. EDGE and Bailando are the state-of-the-art models in dance generation, and GTN-Bailando is an extension of Bailando that incorporates a genre token network to enhance the genre consistency of the generated dances. 
For each approach, we generate 40 dance sequences, each of 20 seconds in duration, using the music from the test set. 

\subsection{Results}
\subsubsection{Quantitative Evaluation}
Following Bailando, we evaluate generated dances in terms of motion quality, motion diversity, and alignment with music rhythms. Specifically, for motion quality, we compute the Fréchet Inception Distances (FID)~\cite{fid} between the generated dances and all the dance sequences of AIST++ on kinetic and geometric features, resulting in two metrics:  $\mathrm{FID}_k$ and $\mathrm{FID}_g$.
For motion diversity, we calculate the average Euclidean distance on kinetic and geometric features, producing two metrics $\mathrm{Div}_k$ and $\mathrm{Div}_g$.
For alignment, we calculate Beat Align Score ($\mathrm{BAS}$)~\cite{bailando} to measure the average distance between each music beat and its nearest dance beat. 

\begin{table*}[ht]
    
    \centering
    \renewcommand{\arraystretch}{1.1}  
    \caption{Evaluation results of different dance generation methods. The results of mean opinion scores (MOS) in user study are shown with 95\% confidence intervals. `w/o frequency' means using VQ-VAE instead of FreqVQ-VAE in ablation study. `w/o MERT(Handcrafted)' means using handcrafted features only and `w/o MERT(Jukebox)' means using Jukebox features instead of MERT features.}
    \label{tab:all_table}
    \resizebox{1\textwidth}{!}{
        \begin{tabular}{cccccccccc}
            \toprule
            \multirow{2}{*}{} & \multicolumn{2}{c}{Motion Quality} & \multicolumn{2}{c}{Motion Diversity} &  & \multicolumn{4}{c}{User Study}\\
            \cmidrule(lr){2-3} \cmidrule(lr){4-5} \cmidrule(lr){7-10} 
            & $\mathrm{FID}_k\downarrow$ & $\mathrm{FID}_g\downarrow$ & $\mathrm{Div}_k\uparrow$ & $\mathrm{Div}_g\uparrow$ & $\mathrm{BAS}\uparrow$
            & OE $\uparrow$ & GM $\uparrow$ & BA $\uparrow$ & DD $\uparrow$\\
            \midrule
            Ground Truth &17.10 &10.60 & 8.19& 7.45& 0.2484 & -- &--& --& --\\
            \hline
            Bailando & 32.10 &9.95 &6.08 &5.70 &0.2299 & 3.28$\pm$0.12  & 3.16$\pm$0.14& 3.22$\pm$0.14 &3.25$\pm$0.13\\
            EDGE &36.23 &23.62 &3.87 &2.65 &0.2430 & 3.52$\pm$0.11  & 3.22$\pm$0.13& 3.53$\pm$0.13 &3.42$\pm$0.12 \\
            GTN-Bailando &29.51 &8.57 &6.15 &6.65 &0.2352 & 3.41$\pm$0.12  & 3.40$\pm$0.13& 3.29$\pm$0.14  &3.27$\pm$0.13\\
            Ours & \textbf{24.25}& \textbf{8.30}&\textbf{7.34} &\textbf{6.91} &\textbf{0.2449} & \textbf{3.70$\pm$0.10}  & \textbf{3.59$\pm$0.14}& \textbf{3.67$\pm$0.13} &\textbf{3.64$\pm$0.12}\\
            \hline
            w/o frequency & 27.80 & 9.84 & 6.73 & 6.05 & 0.2437 & 3.44$\pm$0.12  & 3.48$\pm$0.12& 3.56$\pm$0.12 &3.30$\pm$0.13\\
            w/o MERT(Handcrafted) & 28.49 & 9.83 & 6.28 & 6.28 & 0.2271 & 3.41$\pm$0.09 & 3.19$\pm$0.12& 3.24$\pm$0.12  &3.54$\pm$0.10\\
            w/o MERT(Jukebox) & 28.17 & 9.57 & 7.16 & 6.03 & 0.2207 & 3.54$\pm$0.10 & 3.23$\pm$0.13& 3.31$\pm$0.13  &3.52$\pm$0.11\\
            \bottomrule
        \end{tabular}
    }
\vspace{-0.4cm}
\end{table*}


The results for different methods are presented in Tab.~\ref{tab:all_table}. These results suggest that our method outperforms other methods across all metrics. Specifically, our method achieves the lowest $\mathrm{FID}_k,\mathrm{FID}_g$ and the highest $\mathrm{Div}_k,\mathrm{Div}_g$ and $\mathrm{BAS}$, indicating superior motion quality and motion diversity in the generated dances, and enhanced expressiveness in terms of beat alignment. Notably, the metrics for EDGE differ from those reported in the original paper. 
This discrepancy might arise from the differences in the quantity and duration of the generated dance sequences.

\subsubsection{Qualitative Evaluation}
\begin{figure}[t]
    \centering
    \includegraphics[width=1\linewidth]{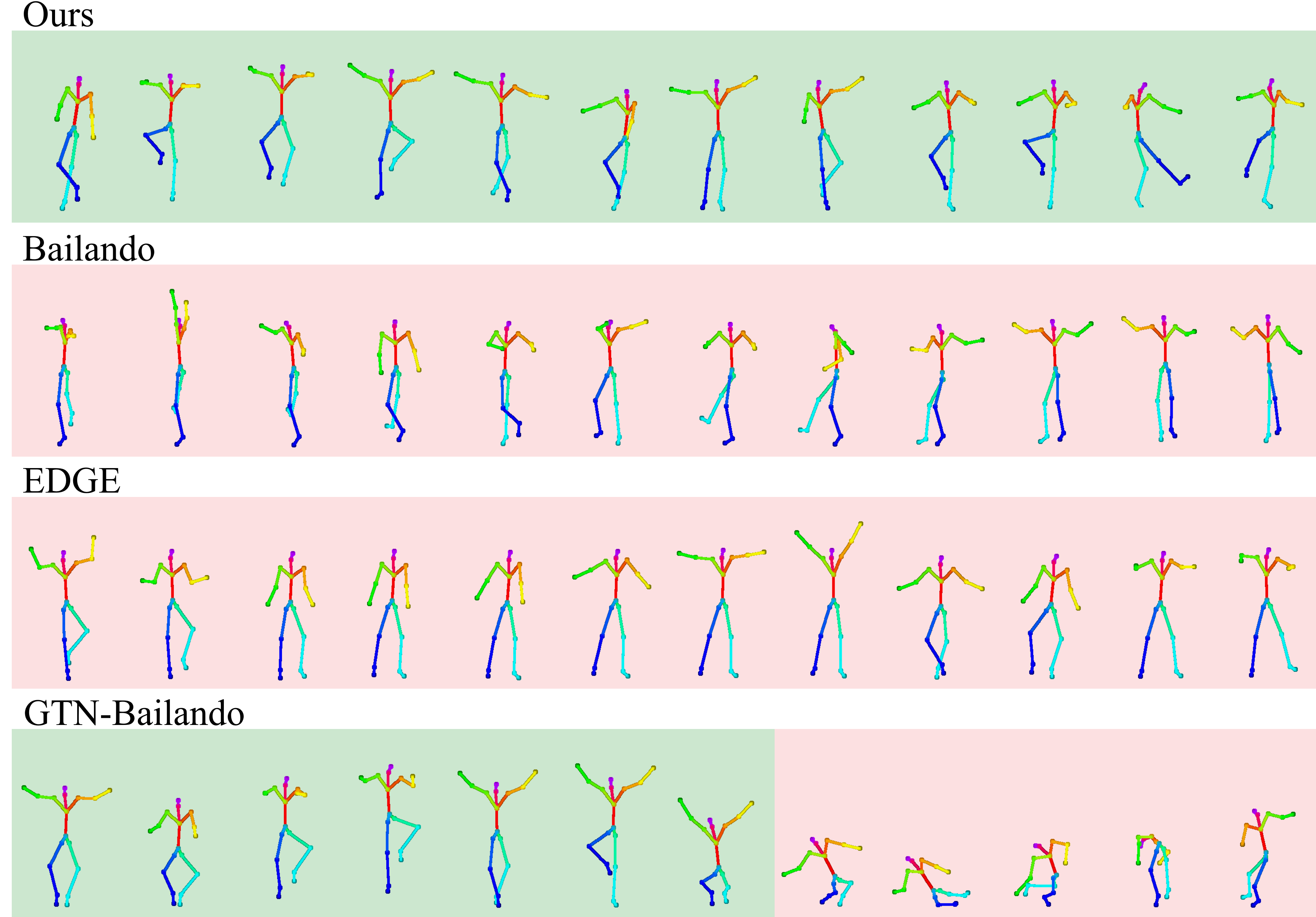} 
    \caption{Dances of genre \textit{Ballet Jazz} generated by different methods. Green background indicates genre matching between dance and music, while red background indicates mismatching. }
    \label{case_study}
    \vspace{-0.45cm}
\end{figure}

Fig.~\ref{case_study} shows the keyframe comparison between each method, taking genre \textit{Ballet Jazz} as an example. Bailando and EDGE generate dances with unpredictable and no-fixed genre because they fail to utilize genre information from music. While GTN-Bailando considers genre information, it still generates genre mismatched motions occasionally, e.g. \textit{footwalks} common in genre \textit{Break} instead of \textit{Ballet Jazz} (see red background in the last row), for weak genre modeling capability of the genre token network trained on limited data. Equipped with pre-trained MERT, which demonstrates strong performance on modeling genre, our method can generate a dance composed of \textit{temps levé}, \textit{pas de bourrée} and \textit{fouetté}, all of which are classic elements of \textit{Ballet Jazz}, revealing better genre matching. More visualization results can be found on our demo page.

\subsubsection{User Study}
Since there are no objective quantitative metrics to directly evaluate dance expressiveness, we conduct a user study to assess the expressiveness of dances generated by different methods. For each method, we select 10 generated dances, 1 for each genre, to conduct the mean opinion scores (MOS) evaluation. We invite 24 participants with prior dance training for this evaluation. Each participant is asked to rate the dance on overall expressiveness (OE) and three specific criteria: genre matching (GM), beat alignment (BA), and dance dynamics (DD). The rating scale ranges from 1 to 5, with 1 being the lowest and 5 the highest, and with a 1-point interval.

The last four columns of Tab.~\ref{tab:all_table} report the results of the MOS evaluation. Our proposed method achieves the best MOS in terms of overall expressiveness, genre matching, beat alignment and dance dynamics. This indicates that our method is capable of generating dances with enhanced expressiveness that align with human subjective perceptions.

\subsubsection{Reconstruction Experiments}

To evaluate the benefits of incorporating frequency information, we compare dances reconstructed by FreqVQ-VAE and VQ-VAE using real dance as a reference. For each joint, we calculate the standard deviation of the speed every two seconds and average all the calculated standard deviations. Tab.~\ref{tab:astd} presents results for the left wrist, chest, and the average across all joints. 
The left wrist represents the joint with notable speed variations, whereas the chest serves as an example joint with few speed variations. 
The findings reveal that when generating identical dance movements, the dances generated by FreqVQ-VAE exhibit a higher speed standard deviation than those by VQ-VAE, aligning more closely with the ground truth. This suggests that FreqVQ-VAE, by integrating frequency information, effectively mitigates the speed homogenization issue observed in dances produced by VQ-VAE and has a wider range of speed variations in the generated dances.

\begin{table}[t]
    \centering
    \renewcommand{\arraystretch}{1.2}  
    \caption{Speed standard deviations for different joints. }
    \label{tab:astd}
    \resizebox{1\linewidth}{!}{
        \begin{tabular}{cccc}
            \toprule
            & Left Wrist $(\times 10^{-2})$ & Chest $(\times 10^{-3})$& All Joints   $(\times 10^{-3})$ \\
            \midrule
            Ground Truth & $1.35$ & $5.12$ & $8.08$\\
            VQ-VAE &$1.23$  & $4.03$& $7.39$\\
            FreqVQ-VAE & $1.32$  & $4.14$ & $7.59$ \\
            
            \bottomrule
        \end{tabular}
    }
    
\vspace{-0.2cm}
\end{table}
\subsubsection{Ablation Study}

    

We conduct ablation studies to further investigate the effectiveness of FreqVQ-VAE and the use of MERT features. As shown in the last rows of Tab.~\ref{tab:all_table}, using VQ-VAE instead of FreqVQ-VAE results in reduced dance dynamics of the generated dances and using handcrafted music features only leads to a decline in genre matching and beat alignment. These results suggest that incorporating frequency information into VQ-VAE enhances dance dynamics, and the integration of genre- and beat-related music style information improves genre matching and beat alignment. Additionally, using Jukebox  instead of MERT to extract music features results in a decrease across all metrics. This demonstrates the effectiveness of using MERT for music feature extraction over Jukebox in dance generation. Moreover, leveraging FreqVQ-VAE or MERT helps to improve the motion quality and motion diversity of the generated dances, as $\mathrm{FID}$ and $\mathrm{Div}$ decrease when they are not used.


\section{CONCLUSION}
\label{sec:typestyle}
In this paper, we propose a novel dance generation method, ExpressiveBailando, to enhance dance expressiveness in terms of genre matching, beat alignment and dance dynamics. By incorporating frequency information into VQ-VAE, our proposed FreqVQ-VAE mitigates the issue of speed homogenization and improves dance dynamics. Meanwhile, genre matching and beat alignment are also enhanced by integrating genre- and beat-related music style information extracted by MERT. Experimental results on AIST++ demonstrate that our method can generate dances with higher expressiveness based on the music.


\textbf{Acknowledgement}: This work is supported by National Natural Science Foundation of China (62076144), Shenzhen Key Laboratory of next generation interactive media innovative technology (ZDSYS20210623092001004) and Shenzhen Science and Technology Program (WDZC20220816140515001, JCYJ202208181010140\\30).

\bibliographystyle{IEEEbib}
\bibliography{strings,refs}

\end{document}